\documentclass[journal]{IEEEtran}

\usepackage{cite}
\usepackage{amsmath,amssymb,amsfonts}
\usepackage{algorithmic}
\usepackage{graphicx}
\usepackage{textcomp}

\usepackage{
amsmath,epsfig}

\usepackage{amsmath,epsfig,amsfonts,amssymb,graphics,theorem,calc,url,bm}
\usepackage{array}
\usepackage{calc}
\usepackage{verbatim}

\usepackage{amsmath}
\usepackage{amsfonts}
\usepackage{mathrsfs}
\usepackage{pifont}
\usepackage{amssymb}
\usepackage{verbatim}
\usepackage{upgreek}
\usepackage{color}
\usepackage{graphicx}
\usepackage{algorithmic}
\usepackage{algorithm}
\usepackage{epsfig}

\usepackage{fancybox}

\usepackage{amsfonts}
\usepackage{mathrsfs}
\usepackage{pifont}

\usepackage{verbatim}
\usepackage{upgreek}
\usepackage{color}

\usepackage{algorithmic}
\usepackage{algorithm}

\usepackage{fancybox}

\newcommand{\bzero}{\mbox{\boldmath{$0$}}}

\newcommand{\bA}{\mbox{\boldmath{$A$}}}
\newcommand{\ba}{\mbox{\boldmath{$a$}}}
\newcommand{\bB}{\mbox{\boldmath{$B$}}}
\newcommand{\bb}{\mbox{\boldmath{$b$}}}

\newcommand{\bI}{\mbox{\boldmath{$I$}}}
\newcommand{\ii}{\mbox{\boldmath $i$}}

\newcommand{\bn}{\mbox{\boldmath{$n$}}}

\newcommand{\bP}{\mbox{\boldmath{$P$}}}

\newcommand{\bQ}{\mbox{\boldmath{$Q$}}}

\newcommand{\bR}{\mbox{\boldmath{$R$}}}

\newcommand{\bs}{\mbox{\boldmath{$s$}}}

\newcommand{\bv}{\mbox{\boldmath{$v$}}}

\newcommand{\bw}{\mbox{\boldmath{$w$}}}
\newcommand{\bX}{\mbox{\boldmath{$X$}}}
\newcommand{\bx}{\mbox{\boldmath{$x$}}}
\newcommand{\bY}{\mbox{\boldmath{$Y$}}}
\newcommand{\by}{\mbox{\boldmath{$y$}}}

\newcommand{\bz}{\mbox{\boldmath{$z$}}}

\newcommand{\bDelta}{\mbox{\boldmath{$\Delta$}}}

\newcommand{\tr}{\mbox{\rm tr}\, }

\newcommand{\rank}{\mbox{\rm Rank}\, }

\newcommand{\ex}{{\bf\sf E}}

\def\proof {{\bf Proof:} }
\def\endproof{\hfill $\Box$}

\newtheorem{theorem}{Theorem}[section]

\newtheorem{proposition}[theorem]{Proposition}

\begin{document}

\title{Enhanced Robust Adaptive Beamforming Designs for General-Rank Signal Model via an Induced Norm of Matrix Errors}

\author{
Yongwei~Huang,~{\it Senior Member, IEEE},~and~~Sergiy~A.~Vorobyov~{\it Fellow, IEEE} 
\thanks{Y. Huang is with School of Information Engineering, Guangdong University of Technology, University Town, Guangzhou, Guangdong 510006, China. Email: ywhuang@gdut.edu.cn.}
\thanks{S. A. Vorobyov is with Department of Signal Processing and Acoustics, School of Electrical Engineering, Aalto University, Konemiehentie 2, 02150 Espoo, Finland. Email: svor@ieee.org.}
}

\maketitle

\begin{abstract}
The robust adaptive beamforming (RAB) problem for general-rank signal model with an uncertainty set defined through a matrix induced norm is considered. The worst-case signal-to-interference-plus-noise ratio (SINR) maximization RAB problem is formulated by decomposing the presumed covariance of the desired signal into a product between a matrix and its Hermitian, and putting an error term into the matrix and its Hermitian. In the literature, the norm of the matrix errors often is the Frobenius norm in the maximization problem. Herein, the closed-form optimal value for a minimization problem of the least-squares residual over the matrix errors with an induced $l_{p,q}$-norm constraint is first derived. Then, the worst-case SINR maximization problem is reformulated into the maximization of the difference between an $l_2$-norm function and a $l_q$-norm function, subject to a convex quadratic constraint. It is shown that for any $q$ in the set of rational numbers greater than or equal to one, the maximization problem can be approximated by a sequence of second-order cone programming (SOCP) problems, with the ascent optimal values. The resultant beamvector for some $q$ in the set, corresponding to the maximal actual array output SINR, is treated as the best candidate such that the RAB design is improved the most. In addition, a generalized RAB problem of maximizing the difference between an $l_p$-norm function and an $l_q$-norm function subject to the convex quadratic constraint is studied, and the actual array output SINR is further enhanced by properly selecting $p$ and $q$. Simulation examples are presented to demonstrate the improved performance of the robust beamformers for certain matrix induced $l_{p,q}$-norms, in terms of the actual array output SINR and the CPU-time for the sequential SOCP approximation algorithm.
\end{abstract}


\IEEEpeerreviewmaketitle

\section{Introduction}
Robust adaptive beamforming (RAB) consists of applying an optimized weight vector (beamvector) to dramatically reduce the sensitivity to any sort of parameter uncertainties and to significantly improve the array output performance. Particularly, the parameters' uncertainty in a RAB design problem can be caused, e.g., by any imperfect knowledge about the source, propagation and sensor array. The array output performance metrics include the output signal-to-interference-plus-noise ratio (SINR) of the beamformer and others \cite{Stoica-Li-book-beamforming,GSSBO09,Voro13survey}.

There are many papers on RAB proposed for the scenario of a general-rank signal model (see \cite{Voro13survey,hvl-tsp-2020} and reference therein), since the desired signal source often can be incoherently scattered (rather than to be a point source corresponding to a rank-one signal model). Often, a RAB design for general-rank signal model addresses a worst-case SINR maximization problem (see, e.g., \cite[problem (46)]{SGLW03}), where the desired signal covariance matrix is not known precisely as well as the sample estimate of the data covariance matrix is inaccurate.
In \cite{SGLW03}, a closed-form solution for the worst-case SINR maximization problem has been proposed; however the positive semidefinite (PSD) assumption on the actual covariance for the desired signal was missing. We have reformulated the maximization problem into a quadratic matrix inequality (QMI) problem via a strong duality of linear conic programming in \cite{hvl-tsp-2020}, incorporating  the PSD assumption, and the global optimality conditions for the QMI problem are established.

On the other hand, the authors of \cite{chen-gershman08} presented an alternative worst-case SINR maximization problem formulation by introducing a matrix decomposition (e.g., spectral type) of the presumed  covariance of the desired signal and putting an error term into both the matrix and its Hermitian obtained from the decomposition. It turned out that the RAB problem is a nonconvex quadratic programming problem, and an approximation algorithm for the quadratic problem has been proposed via solving a semidefinite programming (SDP) problem in each iterative step.
In \cite{chen-gershman11}, a method using a bisection search to solve the RAB problem formulated
in \cite{chen-gershman08} was proposed. In each step of the bisection search, an SDP feasibility problem is solved.  In \cite{KhabVoro-2011Conf,K-V-2013tsp},
the aforementioned RAB problem was viewed as a difference-of-convex (DC) functions optimization problem, and an approximate algorithm has been established, where an SDP problem must be solved in each iterative step. In addition, the authors proved the conditions under which the output by the algorithm is indeed globally optimal. In \cite{HV2018spl}, the RAB problem has been approximated by a sequence of second-order cone programming (SOCP) problems, and an algorithm that exhibited faster run time than the previously proposed SDP based approximation algorithm has been developed.

In this paper, we study the RAB problem, in a form of the worst-case SINR maximization (see, e.g., \cite{K-V-2013tsp}), with the uncertainty set defined via a matrix induced norm of the errors. Then, we can identify a matrix induced $l_{p,q}$-norm such that the RAB problem has a solution improving the actual beamformer output SINR the most. Remark that in contrast to the Frobenius norm of the matrix errors, the introduction of the matrix induced $l_{p,q}$-norm of the errors paves a {\it new way to enhance the array performance}. Specifically, we first derive the closed-form optimal value of a minimization problem of the least-squares residual, where the induced $l_{p,q}$-norm of the matrix errors is bounded by a constant. The optimal value obtained extends the result in \cite{Beck-orl-2009}. Then the optimal value is applied to the worst-case SINR maximization problem with an induced $l_{2,q}$-norm of the matrix errors, and then the RAB problem is recast into a maximization problem with the objective being the difference between an $l_2$-norm function and an $l_q$-norm function, subject to a convex quadratic constraint. The maximization problem can be approximated by a sequence of SOCP problems with nondecreasing optimal values, since the epigraph of an $l_q$-norm function is second-order cone (SOC) representable for any rational number $q$ greater than or equal to one and extended also with the infinity. By comparing the actual array output SINRs computed by the optimal beamvectors for the RAB problems with different values of $q$, we can select the optimal beamvector with a certain matrix induced $l_{2,q}$-norm as a candidate, which gives the maximal actual array output SINR. Then this candidate is the optimal solution. In addition, we consider a generalized RAB problem with an induced $l_{p,q}$-norm of the matrix errors, and the problem again can be solved approximately by a sequential SOCP approximation algorithm for $p\ge1$ and rational number $q\ge1$. Finally, the optimal beamvector corresponding to a matrix induced $l_{p,q}$-norm, which gives the best real array output SINR, is found.

The paper is organized as follows. In Section II, we introduce the signal model and our problem formulation of the RAB problem for general-rank models.
In Section III, we propose the sequential SOCP approximation algorithm for the RAB optimization problem with a rational number $q$ greater than or equal to one. In Section IV, we study generalized RAB problems involving the induced $l_{p,q}$-norm of the matrix errors. Section V includes illustrative numerical examples aiming to show how to select the best matrix induced norm, and Section V draws some concluding remarks.

{\it Notation}: We adopt the notation of using boldface for vectors $\ba$ (lower case), and matrices $\bA$ (upper case). The transpose operator and the conjugate transpose operator are denoted by the symbols $(\cdot)^T$ and $(\cdot)^H$, respectively. The notation $\mbox{tr}(\cdot)$ stands for  the trace of the square matrix argument; $\bI$ and ${\bf 0}$ denote respectively the identity matrix and the matrix (or the row vector or the column vector) with zero entries (their size is determined from the context). The letter $j$ represents the imaginary unit (i.e., $j=\sqrt{-1}$), while the letter $i$ serves as an index. For any complex number $x$, we use $\Re(x)$ and $\Im(x)$ to denote respectively the real and imaginary parts of $x$, $|x|$ and $\mbox{arg}(x)$ represent the modulus and the argument of $x$, and $x^*$ ($\bx^*$ or $\bX^*$) stands for the (component-wise) conjugate of $x$ ($\bx$ or $\bX$).
The Euclidean norm (the Frobenius norm) of the vector $\bx$ (the matrix $\bX$) is denoted by $\|\bx\|$ ($\|\bX\|_F$), while the $q$-norm (for $q\ne2$ and $q\ge0$) of $\bx$ is denoted by $\|\bx\|_q$. The curled inequality symbol $\succeq$ (and its strict form $\succ$) is used to denote generalized inequality: $\bA\succeq\bB$ means that $\bA-\bB$ is a Hermitian positive semidefinite matrix ($\bA\succ\bB$ for positive definiteness).
$\rank(\cdot)$  and $\ex[\cdot]$ represent the rank of a matrix argument and statistical expectation, respectively.
The notation $\lambda_{\max} (\bX)$  stands for the maximal eigenvalue of the square matrix $\bX$. 

\section{Signal Model and Problem Formulation}
The array observation vector is given by
\begin{equation}\label{signal-received}
\by(t)=\bs(t)+\ii(t)+\bn(t)\in{\mathbb{C}^N}
\end{equation}
where $\bs(t)$, $\ii(t)$, and $\bn(t)$ are the statistically
independent components of the desired signal, interference, and array
noise, respectively, and $N$ is the number
of antenna elements of the array. The array output signal can be
written as
\begin{equation}\label{array-output}
x(t)=\bw^H \by(t)
\end{equation}
where $\bw\in{\mathbb{C}^N}$ is the vector of beamformer (termed also beamvector). The beamformer output SINR is obtained by
\begin{equation}\label{SINR}
\mbox{SINR}=\frac{\bw^H\bR_{\rm s} \bw}{\bw^H\bR_{\rm i+n}\bw}
\end{equation}
where the desired signal covariance matrix is $\bR_{\rm s} = \ex[\bs(t) \bs^H(t)]$ and the interference-plus-noise covariance matrix is $\bR_{\rm i+n} = \ex[(\ii(t)+\bn(t))(\ii(t)+\bn(t))^H]$. Note that the SINR \eqref{SINR} is unchanged when the beamvector $\bw$ is replaced with $r e^{j\theta}\bw$, where $r e^{j\theta}$ is any nonzero complex number. Here, matrix $\bR_{\rm s}$ is of general rank, i.e., $\rank(\bR_{\rm s}) \in \{1,\ldots,N\}$. The rank-one $\bR_{\rm s}$ corresponds to the case of a point source or a scattered source in a far field from the array (see \cite{Stoica-Li-book-beamforming, Voro13survey, SGLW03, chen-gershman08}). 

An optimal beamforming problem of maximizing the SINR can be expressed as:
\begin{equation}\label{max-SINR}
\begin{array}[c]{cl}
\underset{\bw\ne\bzero}{\sf{maximize}}  &
\begin{array}[c]{c}\frac{\bw^H\bR_{\rm s}\bw}{\bw^H\bR_{\rm i+n}\bw},
\end{array}
\end{array}
\end{equation}
If $\bR_{\rm s}$ and $\bR_{\rm i+n}$ are known perfectly in some way, the optimal value of \eqref{max-SINR} is the maximal eigenvalue $\lambda_{\max} (\bR^{-1/2}_{\rm i+n} \bR_{\rm s} \bR^{-1/2}_{\rm i+n})$ and the optimal solution is an eigenvector corresponding to this maximal eigenvalue.

However, in practical scenarios, matrix
$\bR_{\rm i+n}$ is not available. As a compromise, the sample covariance matrix for $\bR=\ex[\by(t)\by^H(t)]$:
\begin{equation}\label{sample-cov}
\hat\bR=\frac{1}{T}\sum_{t=1}^T \by(t)\by^H(t)
\end{equation}
is used to replace $\bR_{\rm i+n}$ in the optimal beamforming design problem \eqref{max-SINR}. In \eqref{sample-cov}, $T$ stands for the
number of training snapshots. Of course, there could be substantial difference between $\bR_{\rm i+n}$ and $\hat\bR$.  On the other hand, the  desired  signal covariance matrix $\bR_s$ usually is only imperfectly known; in other words,  there is always a certain mismatch between
the presumed signal covariance matrix $\hat\bR_{\rm s}$ and the actual signal covariance matrix $\bR_{\rm s}$.

If $\hat\bR_{\rm s}$ and $\hat\bR$ are used in \eqref{max-SINR} to compute a beamvector, then the so obtained beamvector often leads to poor performance of the array (in terms of, e.g., the real array output SINR), which means that the beamvector could be useless. Therefore, in order to improve the array performance, RAB techniques have been studied in the last two decades, and there is a number of papers on this subject (see \cite{GSSBO09, Voro13survey, hvl-tsp-2020} and references therein).

Among all existing robust adaptive beamforming problem formulations, the following problem maximizing the worst-case SINR (cf. \cite{Boyd08}) is of great interest:
\begin{equation}\label{max-min-2008}
\begin{array}[c]{cl}
\underset{\bw\ne\bzero}{\sf{maximize}} &\underset{\bDelta_1\in{\cal
B}_1,\bDelta_2\in{\cal B}_2}{\sf{minimize}}
\begin{array}[c]{c}\displaystyle\frac{\bw^H(\hat\bR_s+\bDelta_2)\bw}{\bw^H(\hat\bR+\bDelta_1)\bw}
\end{array}
\end{array}
\end{equation}
where the uncertainty sets ${\cal B}_1$ and ${\cal B}_2$ are given
by
\begin{equation}\label{uncertainty-set-b1}
{\cal B}_1 = \{\bDelta_1 \in \mathbb{C}^{N\times
N}~|~\| \bDelta_1 \|_F\le\gamma,\,\hat\bR + \bDelta_1 \succeq \bzero\},
\end{equation}
and
\begin{equation}\label{uncertainty-set-b2}
{\cal
B}_2 = \{\bDelta_2 \in \mathbb{C}^{L\times
L}~|~\| \bDelta_2 \|_F \le \epsilon,\,\hat\bR_{\rm s} + \bDelta_2 \succeq \bzero\},
\end{equation}
respectively. Here $L$ is the number of sources.

Since $\bDelta_1$ and $\bDelta_2$ are separable in the objective function of \eqref{max-min-2008}, it is not hard to show that \eqref{max-min-2008} is equivalent to the following problem:
\begin{equation}\label{max-min-2008-new-B2}
\begin{array}[c]{cl}
\underset{\bw}{\sf{maximize}} &\underset{\bDelta_2\in{\cal
B}_2}{\sf{min}}
\begin{array}[c]{c} \bw^H(\hat\bR_{\rm s} + \bDelta_2)\bw
\end{array}\\
\sf{subject\;to}               &
 \bw^H(\hat\bR+\gamma\bI)\bw\le1,
\end{array}
\end{equation}
in the sense that if $\bw^\star$ solves \eqref{max-min-2008-new-B2}, then it is optimal for \eqref{max-min-2008} too. It is known that \eqref{max-min-2008-new-B2} can be reformulated into the following QMI problem  (see, e.g., \cite{hvl-tsp-2020}):
\begin{equation}\label{max-min-2008-new-B2-1-2}
\begin{array}[c]{cl}
\underset{\bw,\bY}{\sf{maximize}} &\tr(\hat \bR_{\rm s} \bY)-\epsilon\|\bY\|_F\\
\sf{subject\;to}               &   \bw^H \left( \hat\bR+\gamma\bI \right) \bw\le1,\\
& \bw\bw^H-\bY\succeq\bzero.
\end{array}
\end{equation}
However, how to efficiently obtain a globally optimal solution for \eqref{max-min-2008-new-B2-1-2} remains to be a research topic in general despite there is an approximation algorithm and the global optimality conditions established in \cite{hvl-tsp-2020}.

In many existing works (see, e.g., \cite{chen-gershman08, chen-gershman11, KhabVoro-2011Conf, K-V-2013tsp, HV2018spl}), by incorporating the PSD constraint $\hat\bR_{\rm s} + \bDelta_2\succeq\bzero$ in ${\cal B}_2$, the objective $\bw^H(\hat\bR_{\rm s} + \bDelta_2) \bw$ of \eqref{max-min-2008-new-B2} is replaced by
\begin{equation}\label{new-objective}
\bw^H(\bQ+\bDelta_3)^H(\bQ+\bDelta_3)\bw,
\end{equation}
where $\hat\bR_s=\bQ^H\bQ$, $\bQ\in\mathbb{C}^{M\times N}$, $N\ge
M=\rank(\bR_s)$, and the norm of the distortion $\bDelta_3$ is simply bounded
by $\eta$:
\begin{equation}\label{B3-def-Fnorm}
{\cal B}_3=\{\bDelta_3\in\mathbb{C}^{M\times N}~|~\|\bDelta_3\|_F\le\eta\}.
\end{equation}
With the new objective \eqref{new-objective}, problem \eqref{max-min-2008-new-B2} is reformulated into
\begin{equation}\label{max-min-2008-new-closed-form-obj-0}
\begin{array}[c]{cl}
\underset{\bw}{\sf{maximize}} &\underset{\bDelta_3\in{\cal
B}_3}{\sf{min}}
\begin{array}[c]{c} \bw^H(\bQ+\bDelta_3)^H(\bQ+\bDelta_3)\bw
\end{array}\\
\sf{subject\;to}               &
 \bw^H(\hat\bR+\gamma\bI)\bw\le1.
\end{array}
\end{equation}

It can be shown that RAB problem \eqref{max-min-2008-new-closed-form-obj-0} is equivalent to (see, e.g., \cite{HV2018spl}):
\begin{equation}\label{max-min-2008-new-closed-form-obj}
\begin{array}[c]{cl}
\underset{\bw}{\sf{maximize}} & \|\bQ\bw\|-\eta\|\bw\|\\
\sf{subject\;to}               &
 \bw^H\hat\bR\bw+\gamma\|\bw\|^2\le1,
\end{array}
\end{equation}
due to the fact that
\begin{eqnarray}
& &\min_{\bDelta_3\in{\cal B}_3} \bw^H(\bQ+\bDelta_3)^H(\bQ+\bDelta_3)\bw \\ \label{closed-form-objective}
& &~~~~~~~~~~ = \left( \max\{\|\bQ\bw\|-\eta\|\bw\|,0\} \right)^2
\end{eqnarray}
(see, e.g., \cite{Beck-orl-2009}) and the optimal value of \eqref{max-min-2008-new-closed-form-obj} is nonnegative. In \cite{HV2018spl}, problem \eqref{max-min-2008-new-closed-form-obj} is solved by solving its equivalent problem
\begin{equation}\label{max-min-2008-new-closed-form-obj-equiv}
\begin{array}[c]{cl}
\underset{\bw}{\sf{minimize}} & \bw^H\hat\bR\bw+\gamma\|\bw\|^2 \\
\sf{subject\;to}               &
 \|\bQ\bw\|-\eta\|\bw\|\ge1,
\end{array}
\end{equation}
through designing a sequence of SOCP approximation problems with descent optimal values. It turns out by extensive simulations that the sequential SOCP approximation algorithm is faster than the other existing sequential SDP approximation algorithm for \eqref{max-min-2008-new-closed-form-obj-equiv} in \cite{KhabVoro-2011Conf,K-V-2013tsp}, while both solutions by the two methods are globally optimal.

In robust optimization, one of the major questions is how to define a meaningful uncertainty set (see \cite[page 15]{Nemi-book-2009}). Therefore, there is a new possible dimension to improve the real array output performance by defining a proper error set ${\cal B}_3$. Herein, we try to define ${\cal B}_3$ using different matrix induced norms (instead of using only Frobenius norm), and find the choice of a matrix induced norm such that a solution for \eqref{max-min-2008-new-closed-form-obj-0} leads to the actual array output SINR that is enhanced the most.

\section{Solutions for the RAB Problem with a Matrix Induced Norm of the Errors}
In this section,  we design an approch for solving RAB problem \eqref{max-min-2008-new-closed-form-obj-0} when ${\cal B}_3$ is defined by a certain arbitrary matrix induced norm.

\subsection{The Closed-Form Optimal Value for a Minimization Problem of the Least-Squares Residual}

The matrix induced norm $\|\bDelta\|_{p,q}$ can be expressed as
\begin{equation}\label{matrix-induced-norm-def}
\begin{array}[c]{rcl}
\|\bDelta\|_{p,q}=&\underset{\bv}{\sf{maximize}} & \|\bDelta\bv\|_p \\
 &\sf{subject\;to}               & \|\bv\|_q=1,
\end{array}
\end{equation}
where $\bDelta\in\mathbb{C}^{M\times N}$ and $\bv\in\mathbb{C}^N$ (see, e.g., \cite{Boyd}). For example,  $\|\bDelta\|_{2,2}$ is equivalent to the spectral norm $\|\bDelta\|_2$.

Suppose that
\begin{equation}\label{uncertainty-set-0}
{\cal U}_{p,q}=\{\bDelta~|~\|\bDelta\|_{p,q}\le\eta\},
\end{equation} with $p\ge1$ and $q\ge1$.
Therefore, it is not hard to verify that
\begin{equation}\label{uncertainty-set}
{\cal U}_{p,q}=\{\bDelta~|~\|\bDelta\bv\|_p\le\eta\|\bv\|_q,\,\forall\bv\}.
\end{equation}

Let us consider the minimization problem of the least-squares residual as follows.
\begin{equation}\label{min-residual-prob}
\underset{\bDelta\in{\cal U}_{p,q}}{\sf{minimize}} ~ \|(\bA+\bDelta)\bx-\bb\|_p,
\end{equation}where $\bA\in\mathbb{C}^{M\times N}$, $\bx\in\mathbb{C}^N$ and $\bb\in\mathbb{C}^M$ are parameters.
It is known  (see, e.g., \cite{Beck-orl-2009}) that
\begin{equation}\label{Beck-result-2009}
\underset{\|\bDelta\|_F\le\eta}{\sf{minimize}} ~ \|(\bA+\bDelta)\bx-\bb\| = \max\{\|\bA\bx-\bb\|-\eta\|x\|,0\}.
\end{equation}
Here, we extend the previous result in \eqref{Beck-result-2009} by calculating the closed-form optimal value for \eqref{min-residual-prob}.

\begin{proposition}\label{opt-val-min-residual-prob} It holds that
\begin{equation}\label{min-residual-prob-opt-val-in-proposition}
\underset{\bDelta\in{\cal U}_{p,q}}{\sf{minimize}} ~ \|(\bA+\bDelta)\bx-\bb\|_p = \max\{\|\bA\bx-\bb\|_p-\eta\|\bx\|_q,0\},
\end{equation}where ${\cal U}_{p,q}$ is defined as in \eqref{uncertainty-set-0}.
\end{proposition}
\proof Note that if $\bx=\bzero$, then clearly \eqref{min-residual-prob-opt-val-in-proposition} is true. Suppose that $\bx\ne\bzero$. It follows that
\begin{eqnarray}\nonumber
\|(\bA+\bDelta)\bx-\bb\|_p&\ge&\|\bA\bx-\bb\|_p-\|\bDelta\bx\|_p\\ \label{objective-lower-bound-2}
                          &\ge&\|\bA\bx-\bb\|_p-\eta\|\bx\|_q,
\end{eqnarray}
where the second inequality is due to \eqref{uncertainty-set}. First, we show \eqref{min-residual-prob-opt-val-in-proposition} under the condition that
\begin{equation}\label{assumption-1}
\|\bA\bx-\bb\|_p-\eta\|\bx\|_q\ge0,
\end{equation} and then we prove it under the condition that
\begin{equation}\label{assumption-2}
\|\bA\bx-\bb\|_p-\eta\|\bx\|_q<0.
\end{equation}

Suppose that $\hat\by$ is an optimal solution for the following problem
\begin{equation}\label{max-dual-norm-def}
\begin{array}[c]{cl}
\underset{\by}{\sf{maximize}} & |\by^H\bx| \\
\sf{subject\;to}               & \|\by\|_{p^\prime}=1,
\end{array}
\end{equation}
where $p^\prime\ge1$ fulfills $1/p^\prime+1/q=1$. Therefore, we have $|\hat\by^H\bx|=\|\bx\|_q$ (due to the H${\rm \ddot{o}}$lder inequality for complex vectors (see, e.g., \cite[page 346]{ZhangFZ-book-2009}).

Assume that condition \eqref{assumption-1} is satisfied. Then, we claim that $\|\bA\bx-\bb\|_p>0$ (otherwise, it follows from \eqref{assumption-1} that $\bx=\bzero$, which contradicts the assumption that $\bx\ne\bzero$).

Define a rank-one matrix
\begin{equation}\label{rank-one-Delta}
\hat\bDelta=-\eta e^{-j\arg(\hat\by^H\bx)} \frac{(\bA\bx-\bb)\hat\by^H}{\|\bA\bx-\bb\|_p}.
\end{equation}
It can be easily verified that
\begin{eqnarray}
& &\!\!\!\!\!\!\!\!\!\!\!\!\!\!\!\|(\bA+\hat\bDelta)\bx-\bb\|_p \nonumber\\
&=&\left\|\frac{\bA\bx-\bb}{\|\bA\bx-\bb\|_p}(\|\bA\bx-\bb\|_p-\eta|\hat\by^H\bx|)\right\|_p \nonumber\\ \label{opt-val-for-assum-1}
&=&\|\bA\bx-\bb\|_p-\eta\|\bx\|_q,
\end{eqnarray}
and that for any $\bz\in\mathbb{C}^N$, we have
\begin{eqnarray}
\|\hat\bDelta\bz\|_p&=&\left\|\frac{\bA\bx-\bb}{\|\bA\bx-\bb\|_p}\right\|_p|\eta\hat\by^H\bz| \nonumber\\
&=&\eta|\hat\by^H\bz| \le \eta\|\hat\by\|_{p^\prime}\|\bz\|_q \nonumber\\ \label{opt-soln-for-assum-1}
&=&\eta\|\bz\|_q.
\end{eqnarray}
Therefore, $\hat\bDelta$ is such that both inequalities in \eqref{objective-lower-bound-2} are satisfied as equality, and $\hat\bDelta\in{\cal U}_{p,q}$ (which implies that the optimal value for the minimization problem in \eqref{min-residual-prob-opt-val-in-proposition} is $\|\bA\bx-\bb\|_p-\eta\|\bx\|_q$).

Now, we consider the other condition that  $\|\bA\bx-\bb\|_p-\eta\|\bx\|_q<0$ (i.e., \eqref{assumption-2}). Set
\begin{equation}\label{rank-one-Delta-2}
\hat\bDelta=-e^{-j\arg(\hat\by^H\bx)}\frac{(\bA\bx-\bb)\hat\by^H}{\|\bx\|_q}.
\end{equation}
It is easy to see that the objective function of the minimization problem in \eqref{min-residual-prob-opt-val-in-proposition} at $\hat\bDelta$ is
\begin{eqnarray}
\|(\bA+\hat\bDelta)\bx-\bb\|_p&=&\left\|\bA\bx-\bb-\frac{\bA\bx-\bb}{\|\bx\|_q}|\hat\by^H\bx|\right\|_p\\ \label{opt-val-for-assum-2}
                              &=&0,
\end{eqnarray}
and that for any $\bz\in\mathbb{C}^N$,
\begin{equation}\label{opt-soln-for-assum-2}
\|\hat\bDelta\bz\|_p=\left\|\frac{\bA\bx-\bb}{\|\bx\|_q}\right\|_p |\hat\by^H\bz|<\eta|\hat\by^H\bz|\le\eta\|\bz\|_q,
\end{equation}
where the first inequality holds thanks to \eqref{assumption-2}. Therefore, $\hat\bDelta$ has the property that the objective function of the minimization problem in \eqref{min-residual-prob-opt-val-in-proposition} is equal to zero at $\hat\bDelta$, and $\hat\bDelta\in{\cal U}_{p,q}$ (which means that the optimal value for the minimization problem in \eqref{min-residual-prob-opt-val-in-proposition} is zero since the objective is a norm and zero is attainable).

Thereby, combining \eqref{opt-val-for-assum-1}, \eqref{opt-soln-for-assum-1}, \eqref{opt-val-for-assum-2} and \eqref{opt-soln-for-assum-2} yields \eqref{min-residual-prob-opt-val-in-proposition}, and thus, the proof is complete.
\endproof

{\bf Remark:} It is worth noting our proof for the case when the condition $\|\bA \bx - \bb\|_p - \eta \|\bx\|_q\ge0$ holds is somewhat similar to the proof Theorem~1 in \cite{Bertsimas-ejor-2018}, where the maximization problem
\begin{equation}\label{max-value-closed-form-1}
\underset{\bDelta\in{\cal U}_{p,q}}{\sf{maximize}} ~ \|(\bA+\bDelta)\bx-\bb\|_p = \|\bA\bx-\bb\|_p+\eta\|\bx\|_q,
\end{equation} is studied for real-valued parameters and variables (albeit in a different form). In the complex-valued case, the key difference between the proof herein and the proof in \cite{Bertsimas-ejor-2018} lies in the  maximization problem defined in \eqref{max-dual-norm-def}, the rank-one matrix constructed in \eqref{rank-one-Delta} and the additional proof for the second condition that  $\|\bA\bx-\bb\|_p-\eta\|\bx\|_q<0$.

In particular, when $p=q=2$, \eqref{min-residual-prob-opt-val-in-proposition} reduces to
\begin{equation}\label{min-residual-prob-opt-val-p-q-2}
\max\{\|\bA\bx-\bb\|-\eta\|\bx\|,0\}=\underset{\|\bDelta\|_2\le\eta}{\sf{minimize}} ~ \|(\bA+\bDelta)\bx-\bb\|.
\end{equation}
This is the same as the case where $\|\bDelta\|_2$ is replaced with $\|\bDelta\|_F$ (see \eqref{Beck-result-2009}).

\subsection{Solutions for the RAB Problem with a Matrix Induced Norm of the Errors}
Let us now turn our focus on solving robust adaptive beamforming problem \eqref{max-min-2008-new-closed-form-obj-0}.

Suppose that the error set ${\cal B}_3$ in \eqref{B3-def-Fnorm} is redefined as
\begin{equation}\label{B3-def-2-q-norm}
{\cal B}_3=\{\bDelta\in\mathbb{C}^{M\times N}~|~\|\bDelta\|_{2,q}\le\eta_q\},
\end{equation}
where $q\ge1$. Note that the matrix induced $l_{2,q}$-norm is considered here because the numerator of the objective in \eqref{max-min-2008-new-closed-form-obj-0} is the square of the two-norm term, i.e., $\|(\bQ+\bDelta_3)\bw\|^2$. 

It follows from Proposition \ref{opt-val-min-residual-prob} that the inner minimization problem of \eqref{max-min-2008-new-closed-form-obj-0} has the closed-form optimal value
\begin{equation}\label{opt-val-min-residual-prob-2-p-norm}
\left(\max\{\|\bQ\bw\|-\eta_q\|\bw\|_q,0\}\right)^2.
\end{equation}
Therefore, problem \eqref{max-min-2008-new-closed-form-obj-0} is tantamount to the following problem
\begin{equation}\label{max-min-2008-new-closed-form-obj-2-p-norm-1}
\begin{array}[c]{cl}
\underset{\bw}{\sf{maximize}} & \|\bQ\bw\|-\eta_q\|\bw\|_q\\
\sf{subject\;to}               &
 \bw^H\hat\bR\bw+\gamma\|\bw\|^2\le1,
\end{array}
\end{equation} since the optimal value of it is always nonnegative.

Essentially, problem \eqref{max-min-2008-new-closed-form-obj-2-p-norm-1} is a DC optimization problem (the objective is a difference of two convex functions), and hence it appears hard to solve. The SDP approximation method proposed in \cite{KhabVoro-2011Conf, K-V-2013tsp} is not applicable to solve it since $q$ may not be equal to two as it is necessary for \cite{KhabVoro-2011Conf, K-V-2013tsp}. However, using the fact that the epigraph of the function $\|\bw\|_q$ can be represented as SOC for a rational number $q\ge1$ (see, e.g., \cite{Nemi-book2001}), a sequential SOCP approximation approach 
can be applied to solve problem \eqref{max-min-2008-new-closed-form-obj-2-p-norm-1}.

Specifically, in order to solve problem \eqref{max-min-2008-new-closed-form-obj-2-p-norm-1}, let us rewrite it into
\begin{equation}\label{max-min-2008-new-closed-form-obj-2-p-norm-t}
\begin{array}[c]{cl}
\underset{\bw,\,t}{\sf{maximize}} & t\\
\sf{subject\;to}               & \|\bQ\bw\|\ge t+\eta_q\|\bw\|_q\\
 &\bw^H\hat\bR\bw+\gamma\|\bw\|^2\le1.
\end{array}
\end{equation}
Let $\bw_0$ be an initial point. It can be observed that
\begin{equation}\label{soc-approximation-constraints}
\|\bQ\bw\|\ge\frac{\Re(\bw_0^H\bQ^H\bQ\bw)}{\|\bQ\bw_0\|},
\end{equation}
and, thus, if
\begin{equation}\label{soc-approximation-constraints-1}
\frac{\Re(\bw_0^H\bQ^H\bQ\bw)}{\|\bQ\bw_0\|}\ge t+\eta_q\|\bw\|_q,
\end{equation}
then the first constraint of \eqref{max-min-2008-new-closed-form-obj-2-p-norm-t} is satisfied. Thereby, the following problem is a restriction of \eqref{max-min-2008-new-closed-form-obj-2-p-norm-t}
\begin{equation}\label{max-min-2008-new-closed-form-obj-2-p-norm-t-approx}
\begin{array}[c]{cl}
\underset{\bw,\,t}{\sf{maximize}} & t\\
\sf{subject\;to}               & \frac{\Re(\bw_0^H\bQ^H\bQ\bw)}{\|\bQ\bw_0\|}\ge t+\eta_q\|\bw\|_q\\
 &\bw^H\hat\bR\bw+\gamma\|\bw\|^2\le1,
\end{array}
\end{equation}for any given $\bw_0$ such that $\|\bQ\bw_0\|\ne0$.

It can be easily seen that the first constraint in \eqref{max-min-2008-new-closed-form-obj-2-p-norm-t-approx} (i.e., \eqref{soc-approximation-constraints-1})  can be represented as
\begin{equation}\label{soc-approximation-constraints-1-equiv}
\frac{\Re(\bw_0^H\bQ^H\bQ\bw)}{\|\bQ\bw_0\|}\ge t+\eta_q s,\quad s\ge\|\bw\|_q.
\end{equation}
The first constraint in \eqref{soc-approximation-constraints-1-equiv} is linear and the second one can be represented as a SOC constraint for any rational number $q\ge1$.\footnote{Hereafter, we assume that $q$ is any rational number or infinity by default.}
In fact, when $q=1$, $\|\bw\|_q\le s$ is equivalent to
\begin{equation}\label{L1-norm-equiv}
\sum_{n=1}^N v_n\le s,\quad v_n\ge|w_n|, \quad n=1,\ldots,N,
\end{equation}with auxiliary variables $v_n$. When $q=\infty$, condition $\|\bw\|_q\le s$ amounts to
\begin{equation}\label{L-infty-norm-equiv}
s\ge|w_n|,\quad n=1,\ldots,N.
\end{equation}
When $q=2$, $\|\bw\|_q\le s$ is a conventional SOC constraint.

When $q\notin\{1,2,\infty\}$, constraint $\|\bw\|_q\le s$ is equivalent to the following  conditions: 
\begin{equation}\label{Lp-norm-equiv-2}
\sum_{n=1}^N v_n\le s, \quad |w_n|\le s^{(q-1)/q}v_n^{1/q}, \quad v_n\ge0, \quad \forall n,\,s\ge0,
\end{equation}which are identical to the conditions
\begin{eqnarray}
& & \sum_{n=1}^N v_n\le 1, \quad |w_n|\le\xi_n, \quad \xi_n\le s^{(q-1)/q}v_n^{1/q}, \nonumber \\
& & v_n\ge0,\qquad \forall n,\,s\ge0. \label{Lp-norm-equiv-3}
\end{eqnarray}
It follows from \cite[page 108, example 15]{Nemi-book2001} (although real-valued variable $\bw$ is considered therein) that condition $\xi_n\le s^{(q-1)/q} v_n^{1/q}$ in \eqref{Lp-norm-equiv-3} can be always rewritten into standard SOC constraints for any rational number $q\ge1$, and therefore one can conclude that $\|\bw\|_q\le s$ is SOC representable as long as $q\ge 1$ is a rational number. It is also in line with the statement in \cite[page 109, example 17]{Nemi-book2001}, if $\bw$ is real-value.

{\bf Example~1:} Let us check how to rewrite $\xi_n\le s^{(q-1)/q}v_n^{1/q}$ into standard SOC constraints for $q=4\in [2,\infty)$ and $q=3/2\in[1,2]$. 
When $q=4$, the condition $\xi_n\le s^{(q-1)/q}v_n^{1/q}$ is specified to
$\xi_n\le s^{3/4}v_n^{1/4}$, which can be recast as the following constraints  with auxiliary variables $y_n$: 
\begin{equation}\label{Lp-norm-equiv-q-4-con-1}
\xi_n\le\sqrt{s y_n},\quad y_n\le\sqrt{s v_n},
\end{equation}
which can be reexpressed as SOC constraints
\begin{equation}
\sqrt{\xi_n^2 + \left( \frac{y_n - s}{2} \right)^2} \le \frac{y_n + s}{2}, \quad n=1,\ldots,N,
\end{equation}
and
\begin{equation}
\sqrt{y_n^2 + \left( \frac{v_n - s}{2} \right)^2} \le \frac{v_n + s}{2}, \quad n=1,\ldots,N,
\end{equation}
respectively.

When $q=3/2$, condition $\xi_n\le s^{1/3} v_n^{2/3}$ can be reformulated into $\xi_n\le (\xi_n s v_n^2)^{1/4}$, which is tantamount to the following constraints with auxiliary variables $z_n$:
\begin{equation}\label{Lp-norm-equiv-q-3-2-con-1}
\xi_n\le\sqrt{z_n v_n},\quad z_n\le\sqrt{\xi_n s}.
\end{equation}
The two inequalities in \eqref{Lp-norm-equiv-q-3-2-con-1} can be further rewritten as standard SOC constraints
\begin{equation}
\sqrt{\xi_n^2+ \left( \frac{z_n-v_n}{2} \right)^2} \le \frac{z_n+v_n}{2}, \quad n=1,\ldots,N,
\end{equation}
and
\begin{equation}
\sqrt{z_n^2+ \left( \frac{\xi_n-s}{2} \right)^2}\le \frac{\xi_n+s}{2}, \quad n=1,\ldots,N,
\end{equation}
respectively.

Therefore, problem \eqref{max-min-2008-new-closed-form-obj-2-p-norm-t-approx} indeed is an SOCP for any rational number $q\ge1$. A sequential SOCP approximation algorithm for RAB problem \eqref{max-min-2008-new-closed-form-obj-2-p-norm-t} is summarized as follows.

\begin{algorithm}\caption{An SOCP approximation procedure for solving \eqref{max-min-2008-new-closed-form-obj-2-p-norm-t}}\label{alg-1}
\begin{algorithmic}[1]

\REQUIRE  $\hat\bR$, $\bQ$, $q$, $\gamma$, $\eta$, $\alpha$; 

\ENSURE A solution $\bw^\star$ for problem \eqref{max-min-2008-new-closed-form-obj-2-p-norm-t}; 

\STATE Suppose that $\bw_0$ is an initial feasible point; set $k=0$ and $t_0=\|\bQ\bw_0\|-\eta\|\bw_0\|_q$;

\REPEAT

\STATE solve the SOCP \eqref{max-min-2008-new-closed-form-obj-2-p-norm-t-approx} with $\bw_0$ replaced with $\bw_k$, and obtain solution $(\bw^\star,t^\star)$;

\STATE $\bw_{k+1}=\bw^\star$; $t_{k+1}=t^\star$;

\STATE $k=k+1$;

\UNTIL{$t_{k}-t_{k-1}\le \alpha$}

\end{algorithmic}
\end{algorithm}

It is not hard to show that the optimal values $\{t_k\}$ for \eqref{max-min-2008-new-closed-form-obj-2-p-norm-t-approx} obtained by the algorithm is nondecreasing, namely, $t_1\le t_2\le t_3\le\ldots$, since the optimal solution $(\bw_k,t_k)$ is always feasible for \eqref{max-min-2008-new-closed-form-obj-2-p-norm-t-approx} with $\bw_0$ changed to $\bw_k$. We summarize it as follows. 

\begin{proposition}\label{opt-vals-nondecreasing-1}
Suppose that $\{(\bw_k,t_k)\}$ is the sequence obtained from Algorithm \ref{alg-1}. Then, it holds that $\{t_k\}$ is nondecreasing, namely, $t_1\le t_2\le\cdots\le t_k\le\cdots$.
\end{proposition}

It directly follows from Proposition~\ref{opt-vals-nondecreasing-1} that Algorithm~\ref{alg-1} converges to a locally optimal point as long as the optimal value \eqref{max-min-2008-new-closed-form-obj-2-p-norm-1} is finite.

{\bf Example~2:} Suppose that $q\in\{1,1.5,2,4,\infty\}$ (there are five elements in the set). By calling Algorithm~\ref{alg-1}, we can obtain five optimal beamvectors for problem \eqref{max-min-2008-new-closed-form-obj-2-p-norm-1} corresponding to the five values of $q$, respectively.
Comparing the actual array output SINRs computed by the five optimal beamvectors, we find the best $q$ from the five-element set, in the sense that the resultant beamvector with this $l_q$-norm gives the maximal actual array output SINR, and also in the sense that it is fastest for Algorithm~\ref{alg-1} to output the optimal beamvector for \eqref{max-min-2008-new-closed-form-obj-2-p-norm-t} with the $l_q$-norm. Thus, in the worst-case SINR maximization problem \eqref{max-min-2008-new-closed-form-obj-0}, the uncertainty set of ${\cal B}_3$ defined in \eqref{B3-def-2-q-norm} via the matrix induced $l_{2,q}$-norm is selected as the option that the real array output performance is enhanced the most.

We remark that there is no general rule to select the number $q$ such that an optimal beamvector for \eqref{max-min-2008-new-closed-form-obj-2-p-norm-t} with the $l_q$-norm leads to the highest array output SINR, since the optimal beamvector is data-dependent (it depends on the choices of estimated parameters in problem \eqref{max-min-2008-new-closed-form-obj-2-p-norm-1}). Herein, through the proposed algorithm, we can provide the optimal beamvector for problem \eqref{max-min-2008-new-closed-form-obj-2-p-norm-1} with each number $q$ in a finitely-many-element set, among which the best $q$ is decided according to the corresponding actual beamformer output performance.

\section{Solutions for Generalized Robust Adaptive Beamforming Problems}
In this section, we extend robust beamforming problem \eqref{max-min-2008-new-closed-form-obj-0} to
\begin{equation}\label{max-min-2008-new-closed-form-obj-0-ext}
\begin{array}[c]{cl}
\underset{\bw}{\sf{maximize}} &\underset{\bDelta_3\in{\cal
B}_3}{\sf{min}}
\begin{array}[c]{c} \|(\bQ+\bDelta_3)\bw\|_p
\end{array}\\
\sf{subject\;to}               &
 \bw^H(\hat\bR+\gamma\bI)\bw\le1,
\end{array}
\end{equation}
where the distortion set is redefined as
\begin{equation}\label{B3-def-p-q-norm}
{\cal B}_3=\{\bDelta\in\mathbb{C}^{M\times N}~|~\|\bDelta\|_{p,q}\le\eta_{p,q}\},
\end{equation} and $p\ge1$ and $q\ge1$. Clearly, the problem is equivalent to problem \eqref{max-min-2008-new-closed-form-obj-0} when $p=q=2$. 

Therefore, it follows from Proposition \ref{opt-val-min-residual-prob} that the inner minimization problem in \eqref{max-min-2008-new-closed-form-obj-0-ext} has the optimal value
\[
\max\{\|\bQ\bw\|_p-\eta_{p,q}\|\bw\|_q,0\}.
\]
Hence, problem \eqref{max-min-2008-new-closed-form-obj-0-ext} can be reformulated into
\begin{equation}\label{max-min-2008-new-closed-form-obj-2-ext-pq-norm-t}
\begin{array}[c]{cl}
\underset{\bw,\,t}{\sf{maximize}} & t\\
\sf{subject\;to}               & \|\bQ\bw\|_p\ge t+\eta_{p,q}\|\bw\|_q\\
 &\bw^H\hat\bR\bw+\gamma\|\bw\|^2\le1.
\end{array}
\end{equation} In \eqref{max-min-2008-new-closed-form-obj-2-ext-pq-norm-t}, integers $p\ge1$ and $q\ge1$ are two freedom degrees to improve the performance of the resultant beamformer.

By the H${\rm \ddot{o}}$lder inequality, we obtain that for a fixed vector $\bw_0$,
\begin{equation}\label{soc-approximation-constraints-pq}
\|\bQ\bw\|_p\|\bQ\bw_0\|_{q^\prime}\ge|\bw_0^H\bQ^H\bQ\bw|\ge\Re(\bw_0^H\bQ^H\bQ\bw),
\end{equation} where $q^\prime$ satisfies $1/p+1/q^\prime=1$ (i.e., $q^\prime=p/(p-1)$). In other words,
\begin{equation}\label{soc-approximation-constraints-pq-1}
\|\bQ\bw\|_p\ge\frac{\Re(\bw_0^H\bQ^H\bQ\bw)}{\|\bQ\bw_0\|_{q^\prime}}.
\end{equation}
Similar to problem \eqref{max-min-2008-new-closed-form-obj-2-p-norm-t-approx}, we claim that the following beamforming problem
\begin{equation}\label{max-min-2008-new-closed-form-obj-2-ext-pq-norm-t-approx}
\begin{array}[c]{cl}
\underset{\bw,\,t}{\sf{maximize}} & t\\
\sf{subject\;to}               & \frac{\Re(\bw_0^H\bQ^H\bQ\bw)}{\|\bQ\bw_0\|_{q^\prime}}\ge t+\eta_{p,q}\|\bw\|_q\\
 &\bw^H\hat\bR\bw+\gamma\|\bw\|^2\le1,
\end{array}
\end{equation}for any given $\bw_0$ such that $\|\bQ\bw_0\|_{q^\prime}\ne0$, is a restriction problem for problem \eqref{max-min-2008-new-closed-form-obj-2-ext-pq-norm-t}. Clearly, \eqref{max-min-2008-new-closed-form-obj-2-ext-pq-norm-t-approx} is an SOCP when $q\ge1$ is a rational number, and then \eqref{max-min-2008-new-closed-form-obj-2-ext-pq-norm-t} can be solved as a sequential SOCP approximation, as stated in Algorithm \ref{alg-1} with problem \eqref{max-min-2008-new-closed-form-obj-2-p-norm-t-approx} in step 3 replaced by \eqref{max-min-2008-new-closed-form-obj-2-ext-pq-norm-t-approx}, where $\bw_0$ is substituted with $\bw_k$.

The following proposition is in order.

\begin{proposition}\label{nondecreasing-opt-vals}
Suppose that $\{(\bw_k,t_k)\}$ is a sequence obtained from the sequential SOCP approximation algorithm, and suppose that $\{\|\bQ\bw_k\|/\|\bQ\bw_k\|_{q^\prime}\}$ is nondecreasing. Then, $\{t_k\}$ is a nondecreasing sequence.
\end{proposition}
\proof Assume that at the $k$-th step, we solve \eqref{max-min-2008-new-closed-form-obj-2-ext-pq-norm-t-approx} with $\bw_0$ changed to $\bw_k$, getting an optimal solution $(\bw_{k+1},t_{k+1})$. Therefore, we have
\begin{equation}\label{nondecreasing-opt-vals-eqn-1}
\frac{\Re(\bw_k^H\bQ^H\bQ\bw_{k+1})}{\|\bQ\bw_k\|_{q^\prime}}\ge t_{k+1}+\eta_{p,q}\|\bw_{k+1}\|_q,
\end{equation} and $\bw_{k+1}$ satisfying the second constraint of \eqref{max-min-2008-new-closed-form-obj-2-ext-pq-norm-t-approx}.
Due to the assumption that $\{\|\bQ\bw_k\|/\|\bQ\bw_k\|_{q^\prime}\}$ is nondecreasing, the following inequality holds:
\begin{equation}\label{nondecreasing-opt-vals-eqn-2}
\frac{\|\bQ\bw_{k+1}\|}{\|\bQ\bw_{k+1}\|_{q^\prime}}\ge\frac{\|\bQ\bw_k\|}{\|\bQ\bw_k\|_{q^\prime}},
\end{equation}
which implies that
\begin{equation}\label{nondecreasing-opt-vals-eqn-3}
\frac{\|\bQ\bw_{k+1}\|^2}{\|\bQ\bw_{k+1}\|_{q^\prime}}\ge\frac{\|\bQ\bw_k\|\|\bQ\bw_{k+1}\|}{\|\bQ\bw_k\|_{q^\prime}}\ge\frac{\Re(\bw_k^H\bQ^H\bQ\bw_{k+1})}{\|\bQ\bw_k\|_{q^\prime}},
\end{equation}
This, together with \eqref{nondecreasing-opt-vals-eqn-1}, means that
\begin{equation}\label{nondecreasing-opt-vals-eqn-4}
\frac{\Re(\bw_{k+1}^H\bQ^H\bQ\bw_{k+1})}{\|\bQ\bw_{k+1}\|_{q^\prime}}\ge t_{k+1}+\eta_{p,q}\|\bw_{k+1}\|_q.
\end{equation}
Thereby, $(\bw_{k+1},t_{k+1})$ is feasible for \eqref{max-min-2008-new-closed-form-obj-2-ext-pq-norm-t-approx} with $\bw_0$ changed to $\bw_{k+1}$, and hence the optimal value $t_{k+2}$ is not less than $t_{k+1}$, namely, $t_{k+2}\ge t_{k+1}$. The proof is complete.
\endproof

Remark that when $q^\prime=2$ (i.e., $p=2$), sequence $\{\|\bQ\bw_k\|/\|\bQ\bw_k\|_{q^\prime}\}$ reduces to the all-one sequence, and it follows from the proposition that $t_{k+1}\ge t_{k}$ for $k=1,2,\ldots$, which coincides with the previous statement in Proposition~\ref{opt-vals-nondecreasing-1}.

{\bf Example 3:} Suppose that $p\in\{1,2,\infty\}$ and $q=1$. By the modified algorithm, we can solve problem \eqref{max-min-2008-new-closed-form-obj-2-ext-pq-norm-t} with the three different pairs of $(p,q)$ (i.e., $(p,q)\in\{(1,1), (2,1), (\infty,1)\}$), and get the corresponding three optimal beamvectors. Then we compute the actual array output SINRs by utilizing these three beamvectors and select the best $(p,q)$ among the three pairs, such that the solution for problem \eqref{max-min-2008-new-closed-form-obj-2-ext-pq-norm-t} with this $(p,q)$ leads to the maximal array output SINR. Therefore, in \eqref{max-min-2008-new-closed-form-obj-0-ext}, the error set \eqref{B3-def-p-q-norm} defined via the matrix induced $l_{p,q}$-norm 
is treated as the option such that the real array output SINR is enhanced the most.

In addition, recall that the constraint in \eqref{max-min-2008-new-closed-form-obj-0} or \eqref{max-min-2008-new-closed-form-obj-0-ext} is equivalent to
\begin{equation}\label{constraint-equiv}
\underset{\bDelta_1\in{\cal
B}_1}{\sf{max}}
\begin{array}[c]{c} \bw^H(\hat\bR+\bDelta_1)\bw\le 1.\end{array}
\end{equation}
Let $\hat\bR=\bP^H\bP$, where $\bP\in\mathbb{C}^{M^\prime\times N}$ and $\rank(\hat\bR)=M^\prime$. Then we replace \eqref{constraint-equiv} with
\begin{equation}\label{constraint-equiv-1}
\underset{\bDelta_4\in{\cal
B}_4}{\sf{max}}
\begin{array}[c]{c} \|(\bP+\bDelta_4)\bw\|_{p_1}\le 1,\end{array}
\end{equation}
where
\begin{equation}\label{define-B4-set}
{\cal B}_4=\{\bDelta_4\in\mathbb{C}^{M^\prime\times N}~|~\|\bDelta_4\|_{p_1,q_1}\le\eta_{p_1,q_1}\},
\end{equation}
$p_1\ge1$ and $q_1\ge1$. When $p_1=q_1=2$ particularly, condition \eqref{constraint-equiv-1} reduces to
\begin{equation}\label{constraint-equiv-2}
\underset{\|\bDelta_4\|_2\le\eta_{2,2}}{\sf{max}}
\begin{array}[c]{c} \bw^H(\bP+\bDelta_4)^H(\bP+\bDelta_4)\bw\le 1.\end{array}
\end{equation} Here, the constraint $\|\bDelta_4\|_2\le\eta_{2,2}$ in the maximization problem in \eqref{constraint-equiv-2} can be changed to $\|\bDelta_4\|_F\le\eta_{2,2}$, and the optimal value of the maximization problem keeps unaltered; namely, it is equal to $(\|\bP\bw\|+\eta_{2,2}\|\bw\|)^2$.
 It then follows from \eqref{max-value-closed-form-1} that \eqref{max-min-2008-new-closed-form-obj-0-ext} is generalized further into
\begin{equation}\label{max-min-2008-new-closed-form-obj-2-ext-pq-norm-p1q1-norm}
\begin{array}[c]{cl}
\underset{\bw}{\sf{maximize}} & \|\bQ\bw\|_p-\eta_{p,q}\|\bw\|_q \\
\sf{subject\;to}               & \|\bP\bw\|_{p_1}+\eta_{p_1,q_1}\|\bw\|_{q_1}\le1,
\end{array}
\end{equation}
the epigraph form of which can be expressed as
\begin{equation}\label{max-min-2008-new-closed-form-obj-2-ext-pq-norm-p1q1-norm-rst}
\begin{array}[c]{cl}
\underset{}{\sf{maximize}} & t \\
\sf{subject\;to}               & \|\bQ\bw\|_p\ge t+\eta_{p,q}\|\bw\|_q\\
 & r+\eta_{p_1,q_1}s \le1\\
 & \|\bP\bw\|_{p_1}\le r,\,\|\bw\|_{q_1}\le s.
\end{array}
\end{equation}
Similar to \eqref{max-min-2008-new-closed-form-obj-2-ext-pq-norm-t}, this generalized robust adaptive beamforming problem can be approximated by the following SOCP:
\begin{equation}\label{max-min-2008-new-closed-form-obj-2-ext-pq-norm-p1q1-norm-rst-approx}
\begin{array}[c]{cl}
\underset{}{\sf{maximize}} & t\\
\sf{subject\;to}               & \frac{\Re(\bw_0^H\bQ^H\bQ\bw)}{\|\bQ\bw_0\|_{q^\prime}}\ge t+\eta_{p,q}\|\bw\|_q\\
 & r+\eta_{p_1,q_1}s \le1\\
 & \|\bP\bw\|_{p_1}\le r,\,\|\bw\|_{q_1}\le s,
\end{array}
\end{equation}
where $q\ge1$, $p_1\ge1$ and $q_1\ge1$ are rational numbers, $p^\prime=p/(p-1)$, and $\bw_0$ is any given such that $\|\bQ\bw_0\|_{q^\prime}\ne0$. In other words, \eqref{max-min-2008-new-closed-form-obj-2-ext-pq-norm-p1q1-norm-rst} can be solved by calling Algorithm \ref{alg-1} with problem \eqref{max-min-2008-new-closed-form-obj-2-p-norm-t-approx} in step 3 replaced by \eqref{max-min-2008-new-closed-form-obj-2-ext-pq-norm-p1q1-norm-rst-approx}.


\begin{figure}[h]
	\centering
	\vspace{1.8cm}
	\includegraphics{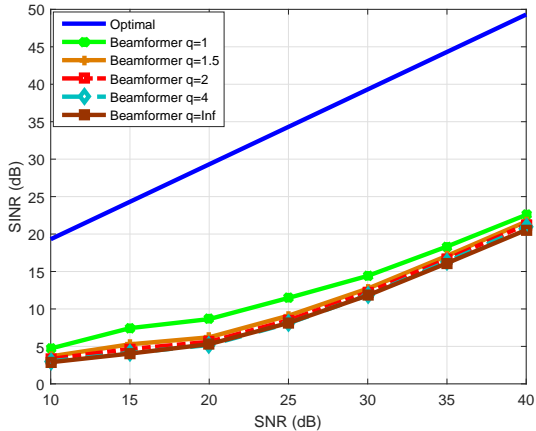}
	\vspace{4.5cm}
	\caption{The beamformer output SINR versus SNR, with INR=10 dB and $T=50$}
	\label{fig-1}
\end{figure}

\section{Simulation Results}
Consider the scenario with a uniform linear array of $N=10$ omnidirectional sensors spaced half a wavelength apart from each other. The additive noise variance in each sensor is set to 0 dB.$~$ There is an interference source with the interference-to-noise ratio (INR) 10 dB impinging on the sensor array. Suppose that both the desired signal and the interference  are locally incoherently scattered sources. The signal of interest and the interference have Gaussian and uniform angular power densities with the central angles 30$^\circ$ and 10$^\circ$, respectively, and  the angular spreads 4$^\circ$ and $10^\circ$, respectively. The presumed signal  of interest is assumed to have Gaussian angular power density with central angle and angular spread 34$^\circ$ and $6^\circ$, respectively. The sample data covariance matrix is estimated with $T=50$ snapshots. The diagonal loading parameter $\gamma=0.005 \|\hat\bR\|_F$ is set. The iteration termination threshold $\alpha$ is equal to $10^{-6}$. All results are averaged over 100 simulation runs.

\subsection{Simulation Example 1: RAB Problem~\label{max-min-2008-new-closed-form-obj-2-p-norm} for $l_{2,q}$-norm }
\begin{figure}[h]
\centering
\vspace{1.8cm}
	\includegraphics{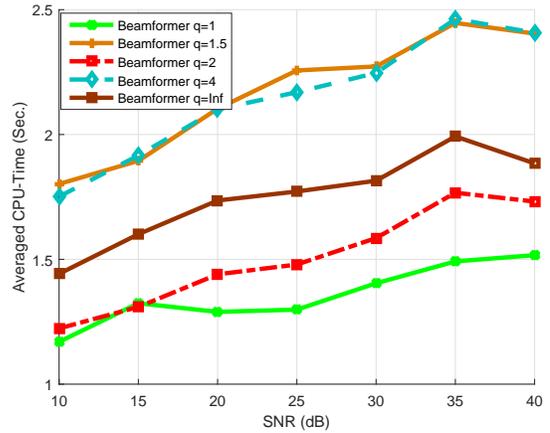}
\vspace{4.6cm}
	\caption{The CPU-time versus SNR, with INR=10 dB and $T=50$}
	\label{fig-2}
\end{figure}

In this example, we examine the array output performance versus signal-to-noise ratio (SNR) obtained by solving RAB problem \eqref{max-min-2008-new-closed-form-obj-2-p-norm-1} with different $q\in\{1,1.5,2,4,\infty\}$, termed as ``Beamformer $q$=1", ``Beamfomer $q$=1.5", ``Beamformer $q$=2", ``Beamformer $q$=4", and ``Beamformer $q$=Inf", respectively. Note that when $q=2$, the beamformer coincides with the one in \cite{HV2018spl}. Let the norm bound $\eta_q=0.05\|\bQ\|_2$ for every $q$ be used.
Fig.~\ref{fig-1} displays the actual array output SINRs versus SNR. As can be seen, the beamformer with $q=1$ has the best performance, which means that the beamformer with $q=1$ is better than the beamformer proposed in \cite{HV2018spl}. Fig.~\ref{fig-2} plots the averaged CPU-time for the proposed algorithm to output a solution with a different $q$, versus SNR. Note that the computer has a processor of Inter Xeon CPU E5-1620 v3 $@$ 3.5GHz
and a 32 GB RAM. We observe that it is fastest to generate the beamformer with $q=1$, which implies that the beamformer outperforms the proposed beamformer in \cite{HV2018spl} in terms of the elapse time of algorithm.

\subsection{Simulation Example 2: RAB Problem~\eqref{max-min-2008-new-closed-form-obj-2-ext-pq-norm-t} for $l_{p,q}$-norm }

\begin{figure}[h]
\centering
\vspace{1.8cm}
	\includegraphics{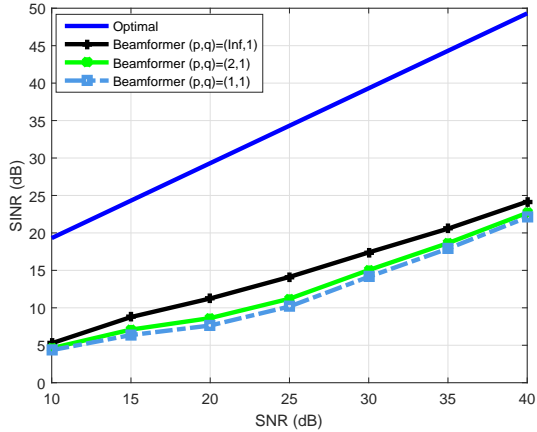}
\vspace{4.5cm}
	\caption{The beamformer output SINR versus SNR, with different $(p,q)$, INR=10 dB and $T=50$}
	\label{fig-3}
\end{figure}

For different pairs of $(p,q)$, we test the array output SINR versus SNR, where a beamvector is obtained by solving problem \eqref{max-min-2008-new-closed-form-obj-2-ext-pq-norm-t} with some $(p,q)$. In particular, $q=1$ is selected since the beamformer with $q=1$ is better than all other beamformers as can be seen in Simulation Example~1, and $p\in\{1,2,\infty\}$. We choose the norm bound $\eta_{p,q}=0.05\|\bQ\|_2$ for every pair $(p,q)$. In Fig.~\ref{fig-3}, it is observed that the beamformer with $(p,q)=(\infty,1)$ outputs the highest SINR in the region of SNR$=[10, 40]$~dB. Note that the beamformer $(p,q)=(2,1)$ coincides with the beamformer with $q=1$ in Simulation Example~1, where $p=2$ and $q=1$ has shown the best performance. This means that the maximal actual array output SINR in Example 1 can be further improved by setting $(p,q)=(\infty,1)$, as observed in this example. 

\begin{figure}[h]
\centering
\vspace{1.8cm}
	\includegraphics{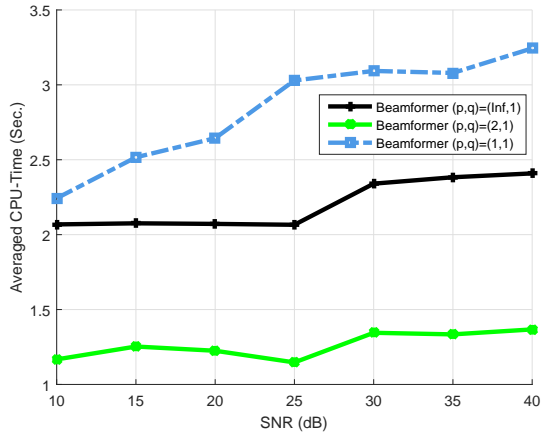}
\vspace{4.6cm}
	\caption{The CPU-time versus SNR, with different $(p,q)$, INR=10 dB and $T=50$}
	\label{fig-4}
\end{figure}

As can be seen in Fig.~\ref{fig-4}, the beamformer with $(p,q)=(2,1)$ is the fastest to converge while the one with $(p,q)=(\infty,1)$ is a bit slower. Thus, there exists a tradeoff. Indeed, the beamformer with $(p,q)=(2,1)$ is fastest, but the output SINR is the second best, while the beamformer with $(p,q) = (\infty, 1)$ is the second fastest, but the output SINR is the best.

\section{Conclusion}
We have considered the RAB problem for general-rank signal model, which has been formulated into a worst-case SINR maximization problem with an uncertainty set defined through a matrix induced norm. We have derived the closed-form optimal value of the minimization problem of the least-squares residual over the matrix errors with an induced norm constraint. Applying the closed-form result, the worst-case SINR maximization problem has been reexpressed as the problem maximizing the difference of a two-norm function and a $l_q$-norm function subject to a convex quadratic constraint. By observing that the epigraph of the $l_q$-norm function with a rational number $q>1$ is SOC-representable, the maximization problem has been solved by a sequential SOCP approximation algorithm. The obtained solutions (beamvectors) corresponding to different values of $q$ are then used compute the actual beamformer output SINRs, and the number $\bar q$ that leads to the maximal beamformer output SINR is selected. In other words, the uncertainty set with matrix induced $l_{2,\bar q}$-norm has been treated as the best choice such that the optimal beamvector for the worst-case SINR maximization problem with the uncertainty set improves the array output performance the most. Apart from that, the matrix induced $l_{p,q}$-norm of the errors has been considered in the worst-case SINR maximization problem, and it has been shown that the problem can be approximated by a sequence of SOCPs. The rational number pair $(\bar p,\bar q)$ which yields the highest array output SINR is chosen. Throughout our simulation examples we have demonstrated how to select the best (in terms of the actual array output SINRs and the CPU-time of the algorithm) uncertainty set defined via a matrix induced norm.

\end{document}